\documentclass[preprint,prb,aps,draft]{revtex4}

\begin{document}

\title{Exposing the spin glass ground state of the non-superconducting La$_{2-x}$Sr$_x$Cu$_{1-y}$Zn$_{y}$O$_4$ high-$T_c$ oxide}

\author{C. Panagopoulos$^1$, A.P. Petrovic$^1$, A.D. Hillier$^2$, J.L.Tallon$^3$, C.A. Scott$^4$, and B.D. Rainford$^4$}
\address{$^1$ Cavendish Laboratory and Interdisciplinary Research Centre in Superconductivity, University of Cambridge, Cambridge CB3 0HE, United Kingdom}
\address{$^2$ Rutherford Appleton Laboratory, Chilton Didcot, Oxfordshire OX11 0QX, United Kingdom}
\address{$^3$ MacDiarmid Institute for Advanced Materials and Nanotechnology, Industrial Research Ltd, P.O. Box 31310, Lower Hutt, New Zealand}
\address{$^4$ Department of Physics and Astronomy, University of Southampton,
Southampton S017 1BJ, United Kingdom}

\date{\today}

\begin{abstract}
We have studied the spin glass behaviour of non-superconducting La$_{2-x}$Sr$_x$Cu$_{0.95}$Zn$_{0.05}$O$_4$ ($x=0.10-0.22$). As in the superconducting analogues of these samples the spin glass transition temperature $T_g$ decreases with increasing $x$, and vanishes at $x=0.19$. A local enhancement in $T_g$ at $x=0.12$ is also observed and attributed to stripe ordering. The disappearance of $T_g$ for $x\geq0.19$ is discussed in terms of a quantum phase transition.
\end{abstract}
\maketitle

%\begin{multicols}{2}

The prototype oxide La$_2$CuO$_4$ is an example of a diluted quantum antiferromagnet in which the Neel temperature decreases with doping and added holes are believed to segregate, at first, into stripes \cite{Sachdev,Kivelson,Zaanen,Smith}. La$_2$CuO$_4$ undergoes an insulator-superfluid transition by hole doping, and this is generally explored by replacing La$^{3+}$ with Sr$^{2+}$ to form La$_{2-x}$Sr$_x$CuO$_4$. Early works indicated that for $x>0.01$ La$_{2-x}$Sr$_x$CuO$_4$ changes from a long-range antiferromagnetic insulator to a spin glass, characterised by a cusp in magnetisation at the glass transition temperature $T_g$ and an associated field hysteresis below \cite{Chou,Wakimoto,Lavrov,Kastner}. The critical exponents and memory effect were found to be in good agreement with conventional spin glass systems, although a crystallographic directional dependence, possibly related to the presence of stripes, was observed. Nevertheless, the onset of spin glass behaviour in all three principle crystallographic directions was shown to be associated with an order parameter and broken symmetry due to frozen spins \cite{SFE1,SFE2}.

Similar magnetisation studies cannot be performed for $x>0.05$ due to masking of the thermodynamic spin glass characteristics by the onset of superconducting diamagnetism. However, other reliable methods such as neutron scattering, nuclear magnetic resonance and muon spin relaxation ($\mu$SR) have been successfully employed to investigate spin glass behaviour across the ($T, x$) phase diagram of HTS \cite{Wakimoto,Kastner,Wakimoto2,Matsuda,Budnick,Harshman,Kiefl,Sternlieb,Niedermayer,Singer1,Singer2,Hunt,Julien1,Julien2,Julien3,CP1,CP2,CP3}. Although $T_g$ obtained from spectroscopic techniques may vary depending on their frequency window there is good agreement with thermodynamic measurements where available. Over the past several years various spectroscopic studies have shown $T_g$ to extend into the superconducting dome of the HTS phase diagram indicating coexistence of superconducting and spin glass order. Furthermore, a comprehensive $\mu$SR study in several HTS showed that $T_g$ ceases to exist at a critical doping $x_c$ $\simeq$0.20 suggesting the presence of a quantum glass transition \cite{CP1,CP2,CP3}.

However, it remains to be confirmed that $x_c$ is in fact robust with respect to the energy window of $\mu$SR. That is, if the absence of dynamical relaxation for $x>x_c$ is independent of the energy scale of the spin fluctuations, hence the presence of an unambiguous quantum transition. To address this question we have exposed the normal state magnetic ground state of La$_{2-x}$Sr$_x$CuO$_4$ across the superconducting dome. Instead of using an applied field, which is beyond experimental capabilities, we use non-magnetic Zn, to suppress superconductivity. Substituting zinc for copper also slows down the spin fluctuations, suppresses long range order, and on the basis of systematic $\mu$SR studies it enhances the muon depolarisation rate at low temperatures and increases $T_g$ \cite{CP2,Kimura}. These effects allow us to eliminate possible masking of a short-range magnetic order due to the $\mu$SR frequency limit. A disadvantage in using zinc is that the exposed ground state is distorted and therefore not pristine. However, this does not affect the aim of the present work since we are mainly interested in the evolution of short-range magnetic order.

The materials studied were La$_{2-x}$Sr$_x$Cu$_{0.95}$Zn$_{0.05}$O$_4$ ($x$=0.10-0.22). Samples were synthesised using solid-state reaction and where necessary followed by quenching and subsequent oxygenation. Spectroscopic, chemical and elemental analyses showed them to be phase pure and stoichiometric. Magnetisation measurements indicated $T_c$=0 in all samples. Zero-field (ZF) and longitudinal-field (LF) $\mu$SR studies were performed at the pulsed muon source, ISIS Facility, Rutherford Appleton Laboratory. In a $\mu$SR experiment, $100\%$ spin-polarised positive muons implanted into a specimen precess in their local magnetic environment. Random spin fluctuations will depolarise the muons provided they do not fluctuate much faster than the muon precession. The muon decays with a life-time $2.2\mu s$, emitting a positron preferentially in the direction of the muon spin at the time of decay. By accumulating time histograms of such positrons one may deduce the muon depolarisation rate as a function of time after implantation. The muon is expected to reside at the most electronegative site of the lattice. As discussed previously in La$_{2-x}$Sr$_x$Cu$_{1-y}$Zn$_{y}$O$_4$ it is the apical O$^{2-}$ nearest to the planes so the results reported here are dominated by the magnetic correlations within the CuO$_2$ planes \cite{CP2,Nachumi}.

Figure 1 (left-hand-side panels) shows the time dependence of the ZF muon asymmetry for La$_{2-x}$Sr$_x$Cu$_{0.95}$Zn$_{0.05}$O$_4$ with $x$=0.10 - 0.18. All samples exhibit glassy behaviour with $x=0.18$ (at $T=0.03$K) being just at the threshold of glassiness. A characteristic signature of a spin-glass within a $\mu$SR spectrum is a rapid initial relaxation followed by a much slower and more gradual long-term relaxation \cite{Uemura}. This can be explained by considering the muon lifetime within the glass: upon insertion, the spin-polarised muons are greeted by a distribution of randomly aligned magnetic fields. Since these fields are quasi-static the detected asymmetry relaxes very quickly. The ensuing longer-term relaxation is due to a small number of low-frequency fluctuations, which may still be present in the CuO$_2$ planes of the samples. Let us note that in a typical spin glass system a single pronounced dip is expected after the sharp drop in asymmetry \cite{Uemura}. However, this behaviour is not universal among systems exhibiting short range ordered magnetism, including HTS, and should not be viewed as the essential condition for characterising spin glass behaviour \cite{Budnick,Harshman,Kiefl,Sternlieb,Niedermayer,CP1,CP2,CP3,Uemura,BDR}. Furthermore, as discussed above, other studies in HTS such as neutron scattering and nuclear magnetic resonance, also show clear evidence for spin glass behaviour. Most importantly however, it is in the temperature region where the ZF $\mu$SR spectrum shows the rapid initial relaxation that thermodynamic measurements exhibit a well-defined field-dependent magnetic transition and associated thermal hysteresis at $T_g$, indicating glassy short-range magnetic ordering with an associated order parameter \cite{Chou,Wakimoto,Lavrov,Kastner}. The right-hand-side panels of Fig. 1 depict typical spectra obtained with an external longitudinal field applied parallel to the initial muon spin polarisation. The field dependence of the spectra indicates that the spins depolarising the muons are static and as the applied field increases the depolarisation decreases and the asymmetry eventually recovers its initial value.

The $x$=0.18 sample displays a much slower relaxation compared with $x<0.18$ but still clearly indicates the presence of a glass at the lowest temperature ($T$=0.03K).  It shows three characteristic indications that it has already entered the glassy phase. Firstly, the spectrum relaxes much quicker than the heavily over-doped samples; secondly, it follows the field dependence of samples $x$=0.10 - 0.16 and thirdly, the rate of change of the gradient in the asymmetry spectrum ($i.e.$ the second derivative) is positive rather than negative. The spectra for $x\geq0.19$ were Gaussian at all temperatures (see e.g. Fig. 2(a)).

The inset for the ZF-plot for $x=0.10$ in Fig. 1 shows ZF-data for $x$=0.10 and 0.12 at short-times. We find that at very small time scales $x=0.10$ and $x=0.12$ (multiplied by 1.3 for clarity) exhibit a faint oscillation which quickly fades as time increases. These oscillations are enhanced at lower temperatures and doping and are successfully fitted by including a precessing component in the Kubo-Toyabe function \cite{Niedermayer,Akoshima,Savici,Adachi}. Oscillations are only observed for $x<0.13$ and $y=0-0.05$ and can only be due to the precession of the muon spins around a stronger non-random internal field caused by the presence of an ordered magnetic state. For example, an antiferromagnetic ground state may be present here. This has also been inferred from neutron scattering experiments for $x$=0.11 \cite{Lake}. We speculate that these oscillations may be related to the presence of stripes \cite{Niedermayer,Akoshima,Savici,Adachi}. At this doping range most HTS show a small drop or a plateau in $T_c$ and in the superfluid density \cite{CP1,CP2,Kivelson2}. The exact reason for this drop is unclear, but it is thought that there is an enhanced tendency towards magnetic ordering at this point due to some peculiarity in the structural orientation of the CuO$_6$ octahedra. This has the effect of creating strongly correlated antiferromagnetic stripe domains. It is these stripes which may provide the spin correlations causing the muon precession. This stripe arrangement is reminiscent of the antiferromagnetic phase present for $x<0.02$ and raises the question as to whether there is a hidden weak antiferromagnetic ground state present in other regions of the phase diagram of La$_{2-x}$Sr$_x$CuO$_4$. Further investigations on this matter are to be carried out.

As the temperature is increased, samples displaying glassy behaviour exhibit a gradual change in the spectra. A typical example, $x=0.10$, is shown in Fig. 2(c). In glassy samples ($x<0.19$), at high temperature the depolarisation is Gaussian and temperature independent, just like $x\geq0.19$, consistent with dipolar interactions between the muons and their near-neighbour nuclear moments.  This was verified by applying a small longitudinal magnetic field (30G) to the $x$=0.14 sample at 50K in order to ascertain the origin of the muon response (Fig. 2(d)).  The field due to the nuclear moments is extremely weak (much weaker than the internal field present for a spin-glass).  Therefore, nuclear decoupling leading to a collapse in the response function should result following application of even a very low field as shown in Fig. 2(d). The Gaussian behaviour at high temperature was verified also for the samples with lower hole-dopings. Furthermore, a 30G field was also applied to the $x$=0.19 sample at $T$=0.03K and 0.75K to check for nuclear decoupling (Fig. 2(b)). The muon response was again observed to collapse, for the same field at both temperatures, showing that there is no evidence for low-energy electron spin fluctuations or onset of glassy behaviour for $x\geq0.19$.

As in earlier works \cite{CP1,CP2,CP3}, two characteristic
temperatures for use in the analysis of the slowing of the spin
fluctuations have been determined. These are the temperature $T_f$
where the spin correlations first enter the $\mu$SR time window,
$i.e.$ where the muon asymmetry first deviates from Gaussian
behaviour and the temperature $T_g$ at which these correlations
freeze into a glassy state thus causing an initial rapid decay in
the asymmetry. The evolution of the muon spin with time ($i.e.$
its relaxation) may be fitted to the form $G_z(t)=A_1 exp( -
\lambda_1 t) +A_2 exp(- (\lambda_2 t)^\beta) +A_3$ where the first
term is the fast relaxation in the glassy state ($i.e.$ $A_1$=0
for $T>T_g$), the second stretched-exponential term describes the
slower relaxation of the dynamical spins and $A_3$ accounts for a
small time-independent background arising from muons stopping in
the silver backing plate of the sample holder. As in some other
spin glass systems, $\beta$ is constant (with the value $\beta$=2)
in the high-temperature Gaussian limit (Fig. 3) \cite{Uemura,BDR}.
The temperature where $\beta$ starts decreasing (Fig. 3) and the
spin lattice relaxation rate $\lambda$ increasing ($e.g.$ Fig. 3
(inset to $x$=0.10)) is regarded as the onset temperature, $T_f$,
at which spin fluctuations slow down sufficiently to enter the
frequency scale of the muon probe ($10^{-10}$s).

Now, to address the question of a possible quantum phase
transition at $x_c$, we concentrate on the freezing temperature,
$T_g$, for which thermodynamic measurements indicate an associated
spin glass order parameter. In ZF-$\mu$SR $T_g$ is identified as
the temperature where $\beta$ falls to 0.5$\pm0.06$ \cite{BDR}.
This root exponential form for the relaxation function is a common
feature of spin glasses. It should be noted that different fitting
procedures for estimating $T_g$ consistently yield similar values
\cite{CP2,Kanigel,Keren}. It is also at $\beta$=0.5 where the spin
lattice relaxation rate, as obtained from the same fits, reaches a
maximum, as shown in the inset to Fig. 3 for $x$=0.10. This
confirms $\mu$SR as a credible technique to accurately identify
and characterise spin-glass behaviour. Figure 3 shows the
variation of $\beta$ with temperature for some of the samples
studied. At first glance we see the onset (where $\beta$ starts
falling) and freezing of spin fluctuations vary in the same manner
as found for pure, 1 and 2 $\%$ Zn doped La$_{2-x}$Sr$_x$CuO$_4$,
$i.e.$, decreasing with increasing doping. Furthermore, for
$x\geq0.19$, there are no changes in the depolarisation function
and $\beta$ is almost temperature independent, consistent with
Gaussian depolarisation. It is clear from the data that changes in
the magnetic ground state occur at $x=0.19$. The main contribution
of the present study is that superconductivity has been fully
suppressed and therefore any masked or hidden magnetism is now
exposed. This was important to demonstrate experimentally since
earlier works left question marks as to whether changes in the
measured magnetic ground state were reflecting actual changes in
ground state of the material, or were an artefact of the presence
of superconductivity.

Values of $T_g$ summarised in Fig. 4 indicate a gradual decrease
of the onset of the spin glass phase with doping. The exception is
the $x=0.12$ sample for which $T_g$ is higher than for $x=0.10$.
We note the positive curvature of $T_g(x)$, exactly as seen
previously for pure, 1 and 2 $\%$ Zn doped La$_{2-x}$Sr$_x$CuO$_4$
and pure Bi$_{2.1}$Sr$_{1.9}$Ca$_{1-x}$Y$_x$Cu$_2$O$_{8+y}$, and
expected in quasi-2D systems like the HTS \cite{CP1,CP2,CP3}. The
increase in $T_g$ in the 1/8 region has been previously discussed
in terms of stripe domains \cite{CP1,CP2,CP3,Akoshima,Adachi}. The
effect of the ordered phase is also evident for $x=0.14$, which
from the trend of $T_g(x)$ indicates the latter is higher than
expected for this doping. In fact the presence of a static stripe
component for $x=0.14$ can be seen in neutron scattering
experiments \cite{Kivelson2,Hirota}. Here the order parameter is
small and is unlikely to be homogeneous which is consistent with
the absence of a well-defined precession frequency in our $\mu$SR
measurements. The spin glass regime vanishes at $x=0.19$ and for
$x\geq 0.19$ we did not observe even an onset ($T_f$) of spin
fluctuations slowing down sufficiently to enter the frequency
scale of the muon probe (Fig. 4). The field dependent studies (see
e.g. Fig. 1) enabled us estimate the internal local field,
$B_{local}$, sensed by the implanted muons \cite{Satooka}. Fig. 4
also includes the doping dependence near $x_c$ of $B_{local}$,
measured at $T=0.03$K. This exhibits a behavior similar to $T_g$
and $T_f$ and disappears at precisely $x=0.19$.

The observation $T_{g,f}(x)$=0 and $B_{local}$=0, at $x\geq 0.19$
$for$ $all$ $Zn$ $concentrations$ indicates that spin glass and
low frequency fluctuations disappear at $x_{c}$ and earlier
results to this effect were not masked by either the frequency
window of the technique or the presence of superconductivity.
Similar conclusions have been reached by recent Cu-nuclear
quadropole resonance studies \cite{Yamagata}. Bearing in mind the
fact that $T_f$ and $T_g \rightarrow 0$ at $x=0.19$ so that the
rate of slowing down actually diverges at this point, the present
results support the existence of a quantum glass transition at
$x_c$. Assuming, based on extrapolation from magnetisation studies
for $x<0.05$ \cite{Chou,Wakimoto,Lavrov,Kastner} and the
systematic trends of the $\mu$SR spectra across the many different
samples we have so-far investigated, the quantum glass transition
is a conventional spin glass transition, in the sense that it has
an associated order parameter and symmetry breaking at $T=0$, then
we may interpret the glass transition at $T=0$ as a quantum
critical point.

We note that the values for $T_g(x)$ for $y=0.05$ are lower than those found in earlier studies for $y=0.02$ even though $T_g(x)$ was found to increase systematically for $x=$0, 0.01 and 0.02 \cite{CP2,CP3}. This presumably occurs because, while Zn slows the spin fluctuations, it also dilutes the spins and at high concentration we see the latter effect. Nevertheless, zinc substitution has undoubtedly exposed the ground state, however distorted it might be, in the samples studied here and allowed us to identify the precise location ($x=0.19$) at which the spin glass disappears. Based also on the systematic tendency of $T_g$ and $T_f$ to vanish at $x=0.19$ it is unlikely $y=0.05$ has suppressed short-range magnetic order only for $x\geq 0.19$. Therefore, we can safely confirm our earlier indications that the magnetic ground state of HTS changes character at $x_c$, independent of the presence or absence of superconductivity.

Other properties also show distinct changes near $x_c$. Recent resistivity studies in $both$ pure and Zn doped high-$T_c$ oxides showed the transition from a pseudogap towards a Fermi liquid phenomenology \cite{Naqib}. Evidence for a change in the ground state has been reported \cite{CP4} for the most fundamental measurable quantity in superconductivity, namely the superfluid density $\rho^{s}$. Both the $ab$-plane and $c$-axis superfluid response in La$_{2-x}$Sr$_x$CuO$_4$ and HgBa$_2$CuO$_{4+\delta}$ were found to remain relatively constant above $x_{c}$ but drop rapidly below $x_{c}$. Furthermore, at 0.19 holes per planar copper atom there is a peak in $\rho^{s}(0)$ for HgBa$_2$CuO$_{4+\delta}$ indicating the strongest superconductivity at the point where $T_g$ vanishes. Also, the doping dependence of the anisotropy in the penetration depth indicates a crossover from two- to three-dimensional transport. These changes in the superconducting ground state, taken together with the disappearance of the spin glass order at the same doping at $T=0$, point consistently to a simultaneous change in the ground state symmetry and superconductivity, as expected in a quantum critical point scenario.

In summary, we have suppressed $T_c$ across the superconducting dome of the La-cuprate family. These samples allowed us to take earlier studies a step further and provide new experimental evidence for absence of short-range magnetism for $x\geq0.19$. The experiments support the presence of a quantum glass transition at $x=0.19$, as also reflected in the disappearance of spin-glass order and the change in the superconducting ground state.

Acknowledgements:
We thank ISIS for the continuing support of this project. C.P. thanks S. Chakravarty and S. Sachdev for useful discussions, and The Royal Society for a University Fellowship. JLT acknowledges financial support from the Marsden Fund.

FIGURE CAPTIONS

FIG. 1.  The panels on the left-hand-side show zero-field $\mu $SR spectra for La$_{2-x}$Sr$_x$Cu$_{0.95}$Zn$_{0.05}$O$_4$ for $x$=0.10 - 0.18. The solid lines are the fits discussed in the text. The right hand panels show the suppression of the glass transition in presence of longitudinal fields. The solid lines are drawn as guide to the eye. The inset shows oscillations seen for $x$=0.10 and 0.12 at low times. The $x=0.12$ data have been shifted for clarity. Data were taken at the temperatures indicated in the respective panels.

FIG.2.  (a) Zero-field $\mu $SR spectra for La$_{2-x}$Sr$_x$Cu$_{0.95}$Zn$_{0.05}$O$_4$ ($x=0.19$)at different temperatures as indicated in the figure. (b) Suppression of the asymmetry of $x=0.19$ at $H$=30G at two different temperatures (see text for details). (c) Zero-field $\mu $SR spectra for $x$=0.10 measured at different temperatures: $T$=1.3K (crosses), 5K (squares), 7K (closed circles), 15K (open circles) and 50K (triangles). (d) Muon asymmetry for $x$=0.14 at high temperature being suppressed by a 30G applied longitudinal field.

FIG. 3.  Typical temperature dependence of the (stretched-exponential) exponent $\beta$ obtained by fitting muon depolarisation data for La$_{2-x}$Sr$_x$Cu$_{0.95}$Zn$_{0.05}$O$_4$ ($x$=0.10, 0.14, 0.16, 0.18, 0.19 and 0.22). The inset to $x$=0.10 shows the temperature dependence of the spin lattice relaxation rate for the same sample.

FIG. 4.  The doping dependence of the temperature $T_g$, below which the spin fluctuations freeze out into a spin glass, for La$_{2-x}$Sr$_x$Cu$_{0.95}$Zn$_{0.05}$O$_4$ ($x$=0.10-0.22). Also shown are data for $T_f$ and $B_{local}$ near $x_c$.

%\end{multicols}

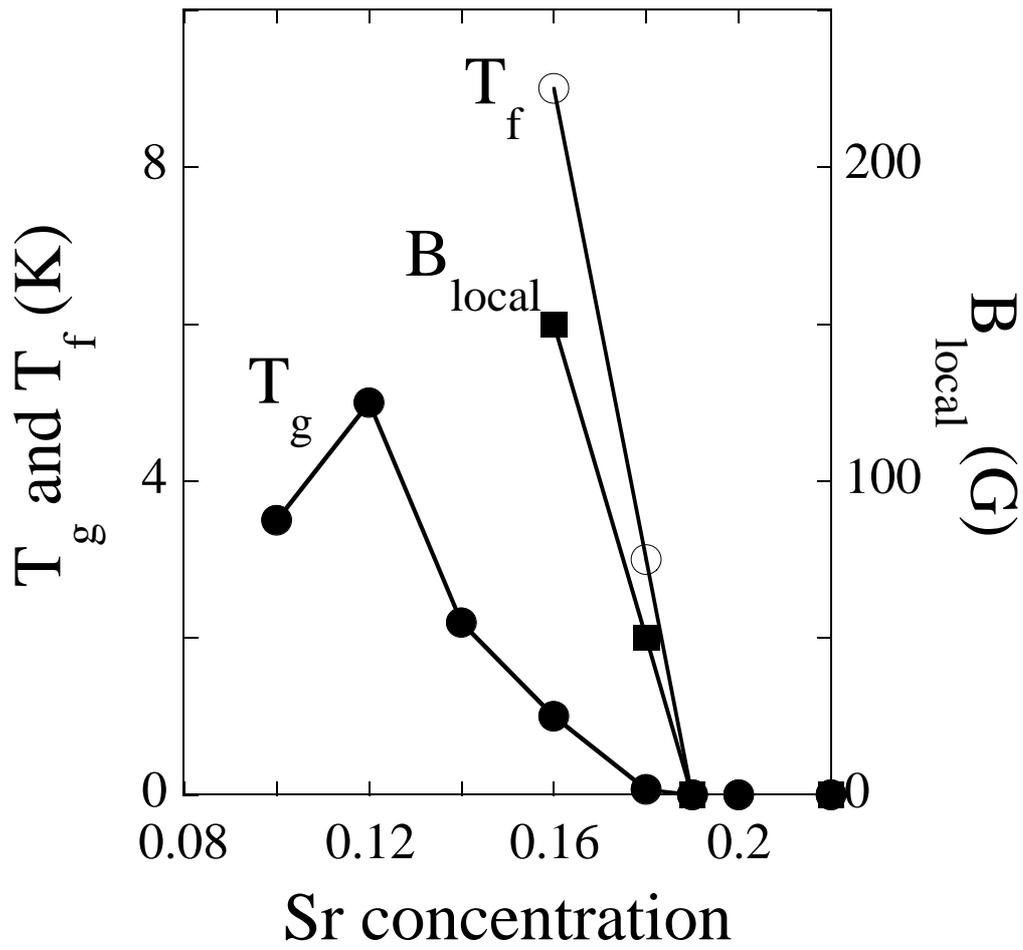
\end{document}